%% file: jss2503.tex
\PassOptionsToPackage{hyphens}{url}
\documentclass[article,shortnames]{jss}
\pdfoutput=1

\usepackage[utf8]{inputenc}
\usepackage{booktabs}
\usepackage{amsmath}
\usepackage{amssymb}
\usepackage{amstext}
\usepackage{epstopdf}
\usepackage{subcaption} 
\usepackage{thumbpdf}
\usepackage{abkuerzungen}

\newtheorem{example}{Example}

\newcommand{\vecmu}{\underline{\mu}}
\newcommand{\vecw}{\underline{w}}
\newcommand{\vecx}{\underline{x}}
\newcommand{\vecz}{\underline{z}}
\newcommand{\matC}{\mathbf{C}}
\newcommand{\matM}{\mathbf{M}}
\newcommand{\matZ}{\mathbf{Z}}
\newcommand{\matB}{\mathbf{B}}
\newcommand{\matSigma}{\boldsymbol{\Sigma}}
\newcommand{\eye}{\mathbf{I}}
\newcommand{\transpose}{\top}
\newcommand{\hermite}{\mathsf{H}}
\newcommand{\complexUnit}{i}

\newcommand{\libname}{libDirectional}

\author{Gerhard Kurz\\Karlsruhe Institute of Technology (KIT)\quad$~$ \And 
        Igor Gilitschenski\\ETH Zurich \AND 
        Florian Pfaff\\Karlsruhe Institute of Technology (KIT)\quad$~$ \And
        Lukas Drude\\University of Paderborn\quad$~$ \AND
        Uwe D. Hanebeck\\Karlsruhe Institute of Technology (KIT) \And
        Reinhold Haeb-Umbach\\University of Paderborn\quad$~$ \AND 
        Roland Y. Siegwart\\ETH Zurich}
\title{Directional Statistics and Filtering\\ Using \libname}

\Plainauthor{Gerhard Kurz, Igor Gilitschenski, Florian Pfaff, Lukas Drude, Uwe D. Hanebeck, Reinhold Haeb-Umbach, Roland Y. Siegwart} 
\Plaintitle{Directional Statistics and Filtering Using \libname} 
\Shorttitle{Directional Statistics and Filtering Using \pkg{\libname}} 

\Abstract{
  In this paper, we present \pkg{\libname}, a \proglang{MATLAB} library for directional statistics and directional estimation. It supports a variety of commonly used distributions on the unit circle, such as the von Mises, wrapped normal, and wrapped Cauchy distributions. Furthermore, various distributions on higher-dimensional manifolds such as the unit hypersphere and the hypertorus are available. Based on these distributions, several recursive filtering algorithms in \pkg{\libname} allow estimation on these manifolds. The functionality is implemented in a clear, well-documented, and object-oriented structure that is both easy to use and easy to extend.
}
\Plainkeywords{recursive filtering, phase estimation, orientation estimation, circle, hypertorus, hypersphere} 

\Address{
	Gerhard Kurz, Florian Pfaff, Uwe D. Hanebeck\\
	Chair for Intelligent Sensor-Actuator-Systems (ISAS) \\
	Institute for Anthropomatics and Robotics\\
	Karlsruhe Institute of Technology (KIT)\\
	Adenauerring 2 \\
	76131 Karlsruhe, Germany \\
	E-mail: \email{gerhard.kurz@kit.edu}, \email{florian.pfaff@kit.edu}, \email{uwe.hanebeck@ieee.org} \\

	Igor Gilitschenski, Roland Y. Siegwart \\
	Autonomous Systems Lab \\
	ETH Zürich \\	
	Leonhardstrasse 21 \\
	8092 Zürich, Switzerland \\
	E-mail: \email{igilitschenski@ethz.ch}, \email{rsiegwart@ethz.ch}\\
	
	Lukas Drude, Reinhold Haeb-Umbach\\
	Department of Communications Engineering \\
	University of Paderborn \\
	Warburger Str. 100 \\
	33098 Paderborn, Germany \\
	E-mail: \email{drude@nt.uni-paderborn.de}, \email{haeb@nt.uni-paderborn.de} 
}

\begin{document}


\newpage

\section{Introduction}
\label{sec:introduction}
\input{introduction.tex}

\newpage 
\section{Related work}
\label{sec:relatedwork}
\input{relatedwork.tex}

\section{Probability distributions}
\label{sec:distributions}
\input{distributions.tex}

\section{Filters}
\label{sec:filters}
\input{filters.tex}

\section{Installation and dependencies}
In this section, we provide a brief explanation of the installation procedure as well as the external libraries \pkg{\libname} depends on.

\subsection{Installation}
The most recent version of \pkg{\libname} can always be obtained from \url{https://github.com/libDirectional}. In order to install the library, the entire \code{lib}-folder including all subdirectories has to be added to \proglang{MATLAB}'s search path. This can be achieved using the \code{startup.m} script. We officially support \proglang{MATLAB} 2014a and later, but most functionality should be available in older versions as well. \pkg{\libname} is platform-independent and runs on the Windows, Linux, and Mac-Versions of \proglang{MATLAB}.

A number of functions of the library have been implemented in \proglang{C++} for performance reasons. These functions can be directly called from \proglang{MATLAB} using the mex-file mechanism. In order to compile the corresponding source code, we provide the \code{compileAll.m} script, which compiles all files that are necessary for \pkg{\libname}, including the external dependencies (see Section~\ref{sec:externals}). The compilation procedure requires a current compiler supported by \proglang{MATLAB} such as Microsoft Visual \proglang{C++} 2013 or later\footnote{There is a bug in the original version of Microsoft Visual \proglang{C++} 2013 that prevents compilation. Installing Microsoft Visual Studio 2013 Update 4 resolves this issue.}, gcc 4.7, or Xcode.

\subsection{Externals}
\label{sec:externals}
To avoid reinventing the wheel, \pkg{\libname} relies on a few libraries written by other authors (see Table~\ref{table:externals}). We include all of these dependencies in the folder \code{externals} to make it easy for the user to run our library. Inclusion of the externals is permitted by their respective licenses and our library \pkg{\libname} itself is licensed under the GPL v3 license.

\begin{table}
	\centering
	\begin{tabular}{lll}
		\toprule
		\bf Name &  \bf Author & \bf License \\
		\midrule
		circVMcdf & Shai Revzen & GPL v3 \\
		Eigen & \cite{eigen2010} & MPL 2 \\ 
		Faddeeva & \cite{johnson2012}  & MIT \\		
		fmath & \cite{fmath2009} & BSD (3-Clause) \\ 
		libBingham & \cite{glover2013lib} & BSD (3-Clause) \\
		mhg & \cite{koev2006} 
		 & GPL v2 or later \\
		Nonlinear Filtering Toolbox & \cite{steinbring2015} &  GPL v3 \\ 
		\bottomrule
	\end{tabular}
	\caption{Dependencies of \pkg{\libname}.}
	\label{table:externals}
\end{table}

\pkg{Eigen} (\cite{eigen2010}) is a \proglang{C++} library for efficient matrix algorithms, which we use for some of our \proglang{C++} implementations. In order to obtain high efficiency, we also use \pkg{fmath} \citep{fmath2009} to calculate exponential functions and logarithms using vector instructions from the SSE/AVX instruction set. 

For the calculation of the normalization constant of the Bingham distribution, we offer several algorithms, one of which is implemented in the \pkg{mhg} library provided by \cite{koev2006}. Furthermore, we use the complex error function required for the multiplication of wrapped normal densities \citep{AES16_Kurz}, which is implemented in the \pkg{Faddeeva} package by \cite{johnson2012}. We also use a modified version of the script \code{circVMcdf} by Shai Revzen, which provides an implementation of the cumulative distribution function of a von Mises distribution as described in~\cite{hill1977}. Moreover, we include some code from \pkg{libBingham} by \cite{glover2013lib}, but this library is not in the ``externals''-folder as only small parts are used and these are directly integrated into the code of \pkg{\libname}.

Finally, we rely on some functions of the \pkg{Nonlinear Filtering Toolbox} for \proglang{MATLAB} by \cite{steinbring2015}. In particular, we use the UKF implementation as well as the deterministic sampling feature for Gaussians from this library. We provide a subset of the \pkg{Nonlinear Filtering Toolbox} within \pkg{\libname} that contains all required functions. 

\section{Conclusion}
In this paper, we have presented \pkg{\libname}, a \proglang{MATLAB} library for directional statistics and directional estimation. As we have shown, this library implements a variety of directional distributions on a number of different manifolds such as the circle, the hypertorus, and the hypersphere. Most distributions offer not only the probability density function but also algorithms for common associated problems such as visualization, parameter estimation, entropy calculation, stochastic sampling, etc. All of these methods are implemented in a clean, object-oriented design that allows the use of analytical solutions whenever possible and provides a transparent fallback to numerical solutions if analytical solutions are unavailable. 

Based on these distributions, a number of different recursive filters are implemented in \pkg{\libname} that can be used for estimation of random variables located on the aforementioned manifolds. These filters not only include many methods based on directional statistics, but also some standard approaches that were modified for the directional setting and that can be used for comparison in order to evaluate the benefits and drawbacks of directional approaches.

We hope that the publication of \pkg{\libname} will make directional statistics and estimation algorithms based thereon available to a wider audience. As the library was designed to be quick to learn and easy to understand, more researches will be able to experiment with these types of methods, to apply them to various problems, and to improve upon them.

\section*{Acknowledgments}

We would like to thank Jannik Steinbring for his helpful advice during the development of \pkg{\libname} as well as for providing prerelease versions of the \pkg{Nonlinear Estimation Toolbox} \cite{steinbring2015}.

%
%
%
%


\bibliography{gk,isaspublikationen_laufend,isaspublikationen,isaspreprints,ld}

\end{document}

%% file: introduction.tex
Directional statistics is a subfield of statistics that deals with quantities defined on manifolds such as the unit circle or the unit hypersphere. Originally mostly developed with geoscientific applications in mind \citep{mardia1981,bingham1974,gaile1980}, directional statistics has gained widespread interest in various areas during the past decades, for example in biology \citep{batschelet1981,mardia2007}, robotics \citep{glover2014icra,feiten2013,markovic2014icra}, machine learning \citep{banerjee2005,gopal2014,diethe2015}, aerospace \citep{horwood2014,darling2015,TAES17_Kurz}, and signal processing \citep{traa2013wrappedkf,azmani2009,drude2014}. A good introduction to the topic can, for example, be found in the book by \cite{mardia1999}.

There are a number of software packages that implement methods stemming from directional statistics (see Section~\ref{sec:relatedwork}). While some of these packages provide good implementations of certain algorithms, most of them are limited to few or just a single probability distribution. Also, usually only a single type of manifold is considered. Moreover, most libraries do not include the ability to perform recursive filtering. To remedy these deficiencies, we present a new library called \pkg{\libname}.

\pkg{\libname} is a library written in \proglang{MATLAB}, a very popular programming language in the engineering community. A few functions are written in \proglang{C} or \proglang{C++} for performance reasons but they can still be conveniently called from \proglang{MATLAB}. The design of the library follows an object-oriented approach, which makes it user-friendly and easily extensible. As the code is thoroughly documented, the library is simple to use, modify, and extend.

We designed \pkg{\libname} with several goals in mind. It is intended to allow the user to easily and quickly experiment with different directional probability densities and filters. Thus, we think it is a valuable tool not only for learning but also for teaching directional statistics. Furthermore, \pkg{\libname} allows rapid prototyping of directional algorithms for a variety of applications. Finally, one of the important goals of the library is to facilitate an easy comparison of different algorithms, for example the quantitative evaluation of a variety of filters.

While it is almost impossible to ensure that any non-trivial software is completely free of bugs, we put a strong emphasis on correctness in the development of \pkg{\libname}. In particular, we implemented a large number of unit tests that can be used to automatically test most of the implemented features and serve as additional usage examples. Aside from the unit tests, we include a lot of assertions \citep{hoare2003} in the code to reduce the risk of problems, for example inadvertently calling certain functions with invalid parameters such as a vector of incorrect dimension or a covariance matrix lacking symmetry or positive definiteness. Even though these assertions introduce some overhead, we decided that early detection of errors and ease of use are more important than speed for \pkg{\libname}. In case the code is used in a real-time environment where speed is critical, it is still possible to remove certain checks to reduce this overhead.


%% file: relatedwork.tex
Over the course of the past two decades, a number of software packages for directional statistics have been developed and published. In this section, we give an overview of the most significant packages.

A lot of the software developed for directional statistics only considers the circular case. An early example is \pkg{CIRCSTAT}, a collection of \proglang{Stata} programs for circular statistics developed by \cite{cox1998}. 
Later, the well-known book on circular statistics by \cite{jammalamadaka2001} was released that includes a floppy disk containing \pkg{CircStats}, a library written in \proglang{S-PLUS} by Ulric Lund. The package \pkg{CircStats} was later ported to \proglang{R} by \cite{agostinelli2012}. An enhanced version of this library was subsequently published under the name \pkg{circular} by \cite{lund2013}. This package is discussed in more detail in the book by \cite{pewsey2013}. Other \proglang{R} libraries such as \pkg{isocir} \citep{barragan2013}, a package for isotonic inference for circular data, and \pkg{NPCirc} \citep{oliveira2014}, a package implementing nonparametric circular regression,  were built on top of \pkg{circular}. It should be noted that \pkg{NPCirc} also includes the ability to perform circular--circular regression (on the torus) and circular--linear regression (on the cylinder).
There are also some packages for other programming languages. A \proglang{MATLAB} toolbox called \pkg{CircStat} (not to be confused with the aforementioned packages \pkg{CIRCSTAT} and \pkg{CircStats}) was published by \cite{berens2009}. As \proglang{MATLAB} is very popular in the engineering community, we have chosen this language for \pkg{\libname} as well. There have also been articles with associated code for circular statistics in \proglang{C++} \citep{krogan2011} and \proglang{Fortran} \citep{allinger2013}. Furthermore, there is a closed-source commercial software called \pkg{Oriana} for circular statistics by \cite{oriana2011}.


While there is quite a lot of software available for the circular case, there are only few packages that deal with hyperspherical data. An early example is \pkg{SPAK} \citep{leong1998}, a package written in  \proglang{MATLAB} that deals  with Kent distributions \citep{kent1982} and offers quite limited functionality. A fairly comprehensive library for the Bingham distribution by the name of \pkg{libBingham} was published by \cite{glover2013lib}. It is written in both \proglang{MATLAB} and \proglang{C}, and can be used from either language. An \proglang{R} package called \pkg{movMF} that deals with mixtures of von Mises--Fisher distributions was later published by \cite{hornik2014}. 

Some software for handling orientations	has also been published. The Bingham-based library \pkg{libBingham} mentioned above is capable of handling rotations using a quaternion representation. Moreover, there is an \proglang{R} package called \pkg{orientlib} by \cite{murdoch2003} and another \proglang{R} package called \pkg{rotations} by \cite{stanfill2014rotations}.



Although the software listed above is very useful for a variety of problems and has successfully been used by many scientists, there are some deficiencies that we seek to address in \pkg{\libname}. Most state-of-the-art software packages are limited to just a single manifold (most frequently the circle) and in some cases to just one particular probability distribution. While this may be fine for scientists only interested in a particular manifold or a particular distribution, in many applications, data on multiple different manifolds is to be considered and more than one probability distribution is of interest. Therefore, we implemented a number of common distributions defined on several manifolds in a unified manner in \pkg{\libname}. The second issue with most existing software is that only very few packages (e.g., \pkg{libBingham}) contain the functionality necessary for recursive estimation. As there is a significant demand for recursive filtering in many applications, e.g., in robotics, autonomous vehicles, aeronautics, etc., we provide several recursive filtering algorithms in \pkg{\libname}.


%% file: distributions.tex
In this section, we introduce the probability distributions implemented in \pkg{\libname}. These distributions can be classified according to the manifold on which they are defined. First, we consider distributions on the unit circle, then probability distributions on the torus, the unit hypersphere, and circular--linear spaces.

\subsection{Circle}
\label{sec:distributions:circle}
In the following, we give an overview of the distributions defined on the unit circle that are implemented in  \pkg{\libname}. In general, we parameterize the unit circle as the half-open interval $[0, 2 \pi)$ while keeping the topology of the unit circle in mind. 

There are several common techniques for deriving circular probability distributions. A widely used method is called \emph{wrapping}. We start with a real random variable $x \sim f(\cdot)$ distributed according to some probability distribution $f(\cdot)$ on $\mathbb{R}$. Now, we consider $x \text{ mod } 2 \pi$, which has the wrapped density 
\begin{align}
	f^\text{wrapped}(t) = \sum_{k=-\infty}^\infty f (t +2 \pi k) \ . \label{eq:wrapping}
\end{align}
This concept has been applied to a number of common distributions, resulting in the wrapped normal (WN), wrapped Cauchy (WC), wrapped exponential (WE), and wrapped Laplace (WL) distributions \citep[Sec.~3.5.7]{jammalamadaka2004,mardia1999}. 
In some cases, the infinite sum in \eqref{eq:wrapping} can be simplified to a closed-form expression (e.g., for the wrapped Cauchy distribution). If a simplification is not possible, it is usually sufficient to consider a small finite number of terms of the series, see \cite{SDF14_Kurz}.
%

Another common concept consists in restricting a linear distribution on $\mathbb{R}^2$ to the unit circle. For example, restricting a two-dimensional Gaussian distribution $\mathcal{N}(\vecx; \vecmu, \matC )$ with
\begin{align*}
	\Vert \vecmu \Vert = 1 \quad \text{and} \quad \matC = \kappa \cdot \begin{bmatrix}1 & 0 \\ 0 & 1 \end{bmatrix}
\end{align*}
to the unit circle, i.e., $\Vert x \Vert=1$, yields an (unnormalized) von Mises (VM) distribution \citep{mises1918}. The von Mises distribution has been further generalized by \cite{gatto2007}.
%

Of course, it is also possible to define distributions directly on the unit circle. For example, the circular uniform distribution and distributions based on Fourier series \citep{willsky1974fourierseries1} belong to this category. Since nontrivial conditions have to be ensured for a Fourier series to be nonnegative \citep{fernandez-duran2007}, we also implemented the option to approximate the square root of the probability density function (pdf) as a Fourier series. By approximating the square root, the pdf values obtained by squaring are always nonnegative and therefore valid according to \cite{Fusion15_Pfaff,Fusion16_Pfaff}. The square root form is used by default but its complexity is hidden from the user. The class \code{FourierDistribution} only requires one additional parameter, the number of coefficients, and can be used like any other distribution in \pkg{\libname}. Furthermore, we provide a class named \code{CircularMixture}, which allows considering mixtures of arbitrary circular distributions, for example von Mises mixtures such as used by \cite{markovic2012}. To represent distributions with a non-standard density (e.g., marginal or conditional densities of certain higher-dimensional distributions), we also offer the class \code{CustomCircularDistribution}.

Aside from the continuous circular distributions, we also consider a discrete circular distribution, which can be thought of as a set of weighted samples. In line with the concept of a Dirac mixture on $\mathbb{R}^n$ \citep{CDC09_HanebeckHuber}, we refer to this distribution as a wrapped Dirac mixture (WD) distribution. For $n$ samples $\beta_1, \dots, \beta_n \in [0,2 \pi)$ with weights $\gamma_1, \dots, \gamma_n > 0$, where $\sum_{j=1}^n \gamma_j = 1$, we write the wrapped Dirac mixture as
\begin{align*}
	\mathcal{WD}(x; \beta_1, \dots, \beta_n, \gamma_1, \dots, \gamma_n) = \sum_{k=-\infty}^\infty \sum_{j=1}^{n} \gamma_j \delta (x + 2\pi k - \beta_j) = \sum_{j=1}^{n} \gamma_j \delta (x - \beta_j) \ , 
\end{align*}
where $x \in [0,2\pi)$ and $\delta(\cdot)$ is the Dirac delta function. Note that this distribution does not have a well-defined probability density function. It is also worth mentioning that this function does not include any wrapping terms because every Dirac component only has ``probability mass'' at a single point \citep[Section~2.2.3~D]{kurz2015}.

\begin{table}[htb]
	\centering
	\small
	\begin{tabular}{ll}
		\toprule
		\bf Class name & \bf Comment \\
		\midrule
		CircularMixture & mixture of arbitrary circular distributions \\ 		
		CircularUniformDistribution & see \cite[Section~2.2.1]{jammalamadaka2001} \\		
		CustomCircularDistribution & distribution with user-specified pdf \\		
		FourierDistribution & see \cite{willsky1974fourierseries1}, \cite{Fusion15_Pfaff}  \\
		GvMDistribution & generalized von Mises, see \cite{gatto2007} \\
		PWCDistribution & piecewise constant (a step function), see \cite{MFI16_Kurz}  \\ 
		VMDistribution & von Mises, see \cite{mises1918} \\
		WCDistribution & wrapped Cauchy, see \cite{jammalamadaka2001}\\ 
		WDDistribution & wrapped Dirac mixture, see \cite{ACC13_Kurz} \\ 
		WEDistribution & wrapped exponential, see \cite{jammalamadaka2004} \\
		WLDistribution & wrapped Laplace, see \cite{jammalamadaka2004} \\
		WNDistribution & wrapped normal, see \cite{schmidt1917} \\
		\bottomrule
	\end{tabular}
	\caption{Probability distributions on the circle.}
	 \label{table:circulardistributions}
\end{table}

All circular distributions implemented in \pkg{\libname} (see Table~\ref{table:circulardistributions}) are derived from an abstract base class called \code{AbstractCircularDistribution}. This base class includes a number of functions that are applicable to all circular distributions and that are independent of the details of the particular distribution. This makes it easy to add implementations of new circular distributions, as a lot of functionality is automatically available once the pdf is defined. In particular, we offer multiple plotting functions to generate different types of visualizations.

\begin{figure}
	\centering
	\begin{subfigure}{.48\textwidth}
	\centering
	\includegraphics[height=5cm]{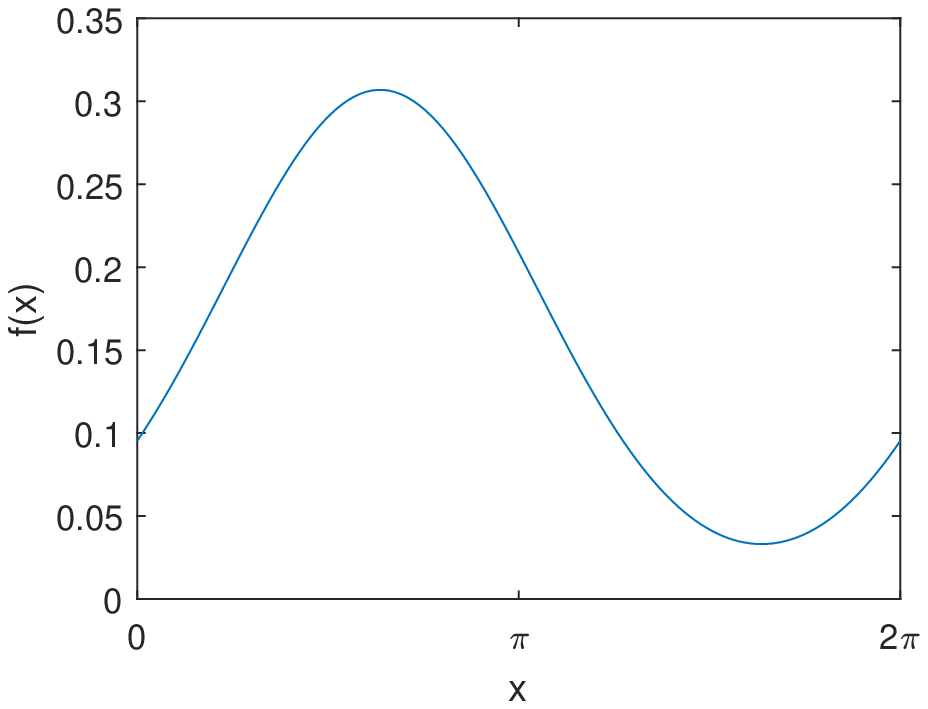}
	\caption{Two-dimensional plot of wrapped normal distribution.}
	\label{fig:wnpdf2d}
	\end{subfigure}
	\quad
	\begin{subfigure}{.48\textwidth}
	\centering
	\includegraphics[height=5cm]{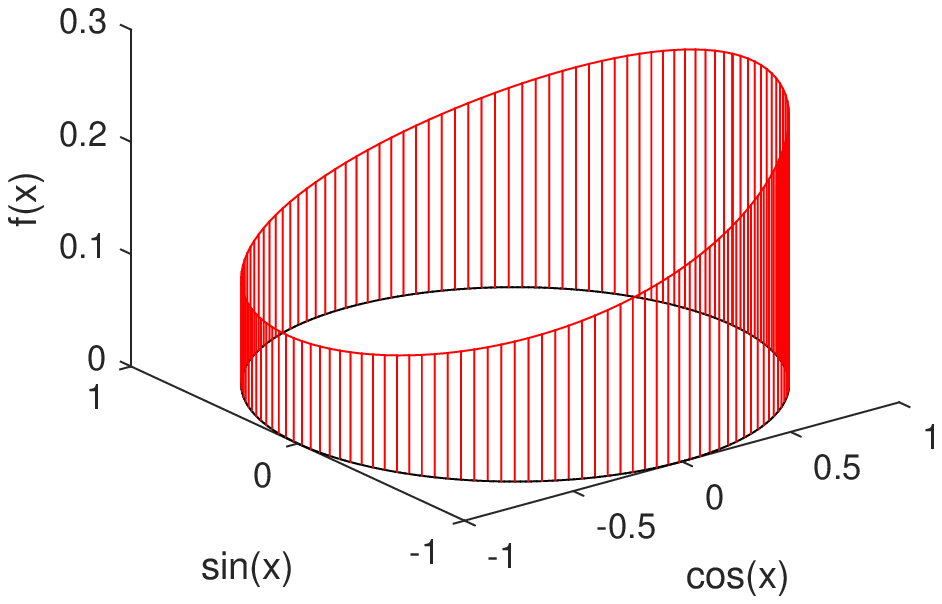}
	\caption{Three-dimensional plot of a von Mises distribution.}
	\label{fig:vmpdf3d}
	\end{subfigure}
	\caption{Visualizations of probability density functions on the unit circle.}
	\label{fig:circularpdf}
\end{figure}

\begin{example}[Plotting probability density functions]
\label{example:plotting}
For example, we can generate a two-dimensional plot of the pdf of a wrapped normal distribution with parameters $\mu = 2$ and $\sigma = 1.3$ simply by typing the following two commands.
\begin{CodeChunk}
\begin{CodeInput}
> wn = WNDistribution(2, 1.3);
> wn.plot();
\end{CodeInput}
\end{CodeChunk}
If we also set the labels and axis using the following code, we obtain the plot depicted in Figure.~\ref{fig:wnpdf2d}.
\begin{CodeChunk}
\begin{CodeInput}
> setupAxisCircular('x');
> xlabel('x'); ylabel('f(x)');
\end{CodeInput}
\end{CodeChunk}
Similarly, we can create plots of other distributions. A three-dimensional plot of the pdf of a von Mises distribution with parameters $\mu = 6$ and $\kappa=0.5$ can be generated using the following code.
\begin{CodeChunk}
\begin{CodeInput}
> vm = VMDistribution(6, 0.5);
> vm.plot3d('color', 'red');
> hold on; vm.plotCircle('color', 'black'); hold off;
> xlabel('cos(x)'); ylabel('sin(x)'); zlabel('f(x)');
\end{CodeInput}
\end{CodeChunk}
The resulting plot is depicted in Figure~\ref{fig:vmpdf3d}. The call to \code{plot3d} visualizes the density itself, whereas the call to \code{plotCircle} creates a circle in the $\cos(x)$--$\sin(x)$--plane.
\end{example}

Also, the abstract base class contains a number of numerical methods to calculate the entropy, trigonometric moments, integrals of the pdf, etc. For most numerical methods (designated by the suffix \code{Numerical}), there are counterparts without this suffix that can be overridden by child classes to provide an analytical implementation. If the child class does not provide an analytical version of the algorithm, the numerical method is used as a fallback.

\begin{example}[Numerical and analytical calculation]
\label{example:numericalanalytical}
Let us consider the wrapped normal distribution defined in Example~\ref{example:plotting} again. Suppose we want to calculate the first trigonometric moment of this distribution, i.e., $\E(\exp(\complexUnit x))$, where $\E(\cdot)$ is the expected value. For this purpose, we simply call the corresponding function:
\begin{CodeChunk}
\begin{CodeInput}
> wn.trigonometricMoment(1)
\end{CodeInput}
\begin{CodeOutput}
ans =
  -0.1788 + 0.3906i
\end{CodeOutput}
\end{CodeChunk}
In the case of the wrapped normal distribution, \code{trigonometricMoment} is a function inside the class \code{WNDistribution} that implements an analytic calculation of the trigonometric moment. If no analytic solution was implemented, the function \code{trigonometricMoment} in the base class \code{AbstractCircularDistribution} would have automatically fallen back to an algorithm based on numerical integration. Even though an analytical solution is available for the wrapped normal distribution, we can still call the numerical algorithm as follows.
\begin{CodeChunk}
\begin{CodeInput}
> wn.trigonometricMomentNumerical(1)
\end{CodeInput}
\begin{CodeOutput}
ans =
  -0.1788 + 0.3906i
\end{CodeOutput}
\end{CodeChunk}
This can, for example, be used to compare the numerical and analytical results in order to validate the correctness of the analytical implementation. In this case, both results match up to the displayed number of digits, but in certain cases, analytical and numerical solutions may differ more significantly. Also, the numerical computation is typically slower, in some cases by several orders of magnitude.
\end{example}

A variety of methods are implemented for some or all circular distributions, for example the probability density function, the cumulative distribution function, the 
circular mean and variance, trigonometric moments, entropy, stochastic sampling, parameter estimation, conversions using trigonometric moment-matching, etc. These methods are thoroughly documented within the code, so we do not go into detail about them here.

Furthermore, we offer convolution and multiplication operations of circular probability densities for some distributions. These operations are required for the circular filtering algorithms discussed in Section~\ref{sec:filters}. For example, we implement the approximations for the von Mises distribution discussed in \cite{azmani2009}, and the approximations for the WN distribution discussed in \cite{AES16_Kurz} and \cite{traa2013wrappedkf}.

One feature, however, deserves a more thorough discussion as many readers may not be familiar with it. In \pkg{\libname}, we have implemented several algorithms for deterministic sampling, a concept where samples are drawn from a distribution deterministically rather than stochastically. These samples are then represented using a (wrapped) Dirac mixture. The advantage of deterministic approaches is that the samples can be placed at representative positions to achieve a good representation of the true density with very few samples. This can, for example, be achieved by performing moment matching. Algorithms of this type have previously been used in Gaussian filters such as the unscented Kalman filter (UKF) by \cite{julier2004} or the smart sampling Kalman filter (S$^2$KF) by \cite{JAIF16_Symmetric_S2KF_Steinbring}. We have proposed deterministic sampling schemes based on approximation of the first trigonometric moment and the first two trigonometric moments in \cite{Fusion14_KurzGilitschenski}. Furthermore, we have proposed approximations using quantization \citep{Fusion16_Gilitschenski}, superposition of moment-based samples, and a binary tree approximation \citep{JAIF16_Kurz}. These methods are implemented in \pkg{\libname} and their use is demonstrated in the following example.

\begin{figure}
	\centering
	\includegraphics[width=8cm]{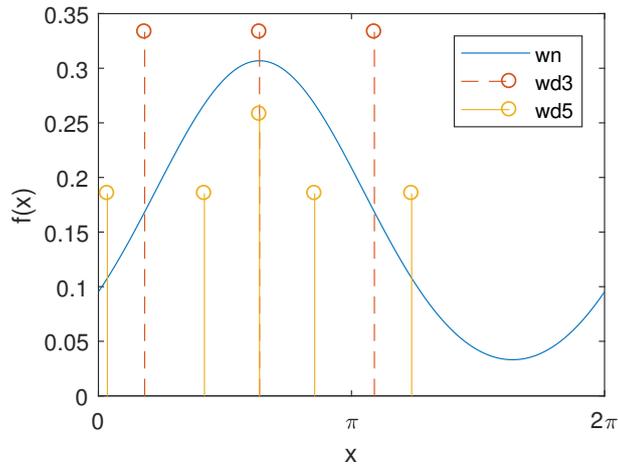}
	\caption{Example for deterministic sampling. The wrapped normal distribution is approximated with either three or five samples. Note that the height of the Dirac delta components is used to represent their weight.}
	\label{fig:deterministicsampling}
\end{figure}

\begin{example}[Deterministic sampling]
\label{example:detsampling}
Once again, we consider the wrapped normal distribution defined in Example~\ref{example:plotting}. Suppose we want to approximate this distribution using a wrapped Dirac mixture with three components, i.e., with three samples on the unit circle. According to \cite{Fusion14_KurzGilitschenski}, the approximation can preserve the first trigonometric moment. We can easily achieve this using the following command.
\begin{CodeChunk}
\begin{CodeInput}
> wd3 = wn.toDirac3()
\end{CodeInput}
\begin{CodeOutput}
wd3 = 
  WDDistribution with properties:
    dim: 1
      d: [0.5740 2 3.4260]
      w: [0.3333 0.3333 0.3333]
\end{CodeOutput}
\end{CodeChunk}
The row vectors $d$ and $w$ correspond to the positions and weights of the Dirac components, respectively. It can be seen that the resulting wrapped Dirac mixture is evenly weighted. We can verify that the first moment is indeed preserved, similar to Example~\ref{example:numericalanalytical}.
\begin{CodeChunk}
\begin{CodeInput}
> mwd = wd3.trigonometricMoment(1), mwn = wn.trigonometricMoment(1)
\end{CodeInput}
\begin{CodeOutput}
mwd =
  -0.1788 + 0.3906i
mwn =
  -0.1788 + 0.3906i  
\end{CodeOutput}
\end{CodeChunk}
An approximation with five components based on the first two trigonometric moments is also possible.
\begin{CodeChunk}
\begin{CodeInput}
> wd5 = wn.toDirac5()
\end{CodeInput}
\begin{CodeOutput}
wd5 = 
  WDDistribution with properties:
    dim: 1
      d: [0.1113 3.8887 1.3156 2.6844 2]
      w: [0.1855 0.1855 0.1855 0.1855 0.2581]
\end{CodeOutput}
\end{CodeChunk}
In this case, the mixture components are not evenly weighted. We can plot the resulting approximations using the following statements.
\begin{CodeChunk}
\begin{CodeInput}
> wn.plot(); hold on; wd3.plot('--'); wd5.plot(); hold off;
> setupAxisCircular('x');
> xlabel('x'); ylabel('f(x)'); legend('wn', 'wd3', 'wd5');
\end{CodeInput}
\end{CodeChunk}
The result is depicted in Figure~\ref{fig:deterministicsampling}.
\end{example}

\subsection{Torus and hypertorus}
Another interesting manifold is the torus as it can be used to represent two (possibly correlated) angular quantities. We parameterize the torus as $[0,2\pi)^2$, i.e., the Cartesian product of two circles.

On the torus, we consider several different distributions, the bivariate wrapped normal distribution, two versions of the bivariate von Mises distribution, the bivariate wrapped Dirac mixture, and a distribution based on a two-dimensional Fourier series (see Table~\ref{table:torusdistributions}). As their names suggest, these distributions constitute bivariate generalizations of the wrapped normal distribution, the von Mises distribution, the wrapped Dirac mixture, and the Fourier distribution, respectively. In the case of the wrapped normal distribution, the generalization to a higher number of dimensions is straightforward \citep[Example 7.3]{johnson1977}. A bivariate wrapped normal distribution arises when a random variable distributed according to a bivariate normal distribution is wrapped in both dimensions. The bivariate wrapped Dirac distribution is obtained analogously. For the bivariate Fourier distribution, a two-dimensional Fourier series is used to approximate the density, or the square root thereof. The bivariate von Mises distribution is more tricky as there are several different, non-equivalent definitions, some of which are discussed in \cite{mardia2007}. In \pkg{\libname}, we chose to implement the sine version as well as the matrix version of the bivariate von Mises. The sine version has the advantage that it has been more thoroughly investigated than its alternatives and a number of its properties are known, e.g., a series representation for its normalization constant \citep{singh2002}. On the other hand, the matrix version has the significant advantage that it is closed under multiplication \citep{MFI15_Kurz}.

\begin{table}[htb]
	\centering
	\setlength{\tabcolsep}{3pt}
	\begin{tabular}{ll}
		\toprule
		\bf Class name & \bf Comment \\
		\midrule
		CustomToroidalDistribution & allows any user-specified pdf \\		
		ToroidalMixture &  mixture of arbitrary toroidal distributions \\ 		
		ToroidalFourierDistribution & bivariate Fourier distribution \citep{JAIF16_Pfaff}\\		
		ToroidalUniformDistribution & uniform distribution on the torus\\		
		ToroidalVMMatrixDistribution & bivariate VM, matrix version~\citep{MFI15_Kurz} \\
		ToroidalVMSineDistribution & bivariate VM, sine version~\citep{singh2002} \\
		ToroidalWDDistribution & bivariate wrapped Dirac mixture \citep{CDC14_Kurz} \\
		ToroidalWNDistribution & bivariate WN \citep{CDC14_Kurz,CDC15_Kurz} \\
		\bottomrule
	\end{tabular}
	\caption{Probability distributions on the torus.}
	\label{table:torusdistributions}
\end{table}

The overall design of toroidal distributions in \pkg{\libname} is similar to the circular distributions discussed above. Once again, there is an abstract base class from which all toroidal distributions inherit. Its name is \code{AbstractToroidalDistribution} and it implements a number of methods that are independent of the particular toroidal distribution. These methods can be overridden by the child classes if analytical solutions are available.


One of the key problems when dealing with toroidal distributions is the question of how to quantify correlation. Over the past decades, a number of different correlation coefficients have been proposed, and we have decided to implement several of them in \pkg{\libname}, namely the correlation coefficients by \cite{johnson1977}, \cite{jupp1980}, and \cite{jammalamadaka1988}. More generally, for a toroidal random vector $[x_1, x_2]^\transpose$, it is of interest to consider 
\begin{align}
	\label{eq:toroidalmoments}
	\tilde{\vecmu} = \E \left( \begin{bmatrix} \cos(x_1) \\ \sin(x_1) \\ \cos(x_2) \\ \sin(x_2) \end{bmatrix} \right) , \
	\tilde{\matC} = \E \left( \left( \begin{bmatrix} \cos(x_1) \\ \sin(x_1) \\ \cos(x_2) \\ \sin(x_2) \end{bmatrix} - \tilde{\vecmu} \right) \cdot \left( \begin{bmatrix} \cos(x_1) \\ \sin(x_1) \\ \cos(x_2) \\ \sin(x_2) \end{bmatrix} - \tilde{ \vecmu} \right)^\transpose \right) \ ,
\end{align}
which we have implemented under the name \code{mean4D} and \code{covariance4D}, respectively. These values can, for example, be used to determine the circular mean in each dimension as well as certain circular--circular correlation coefficients.

\begin{figure}
	\centering
	\begin{subfigure}{.48\textwidth}
		\centering
         \includegraphics[width=6cm]{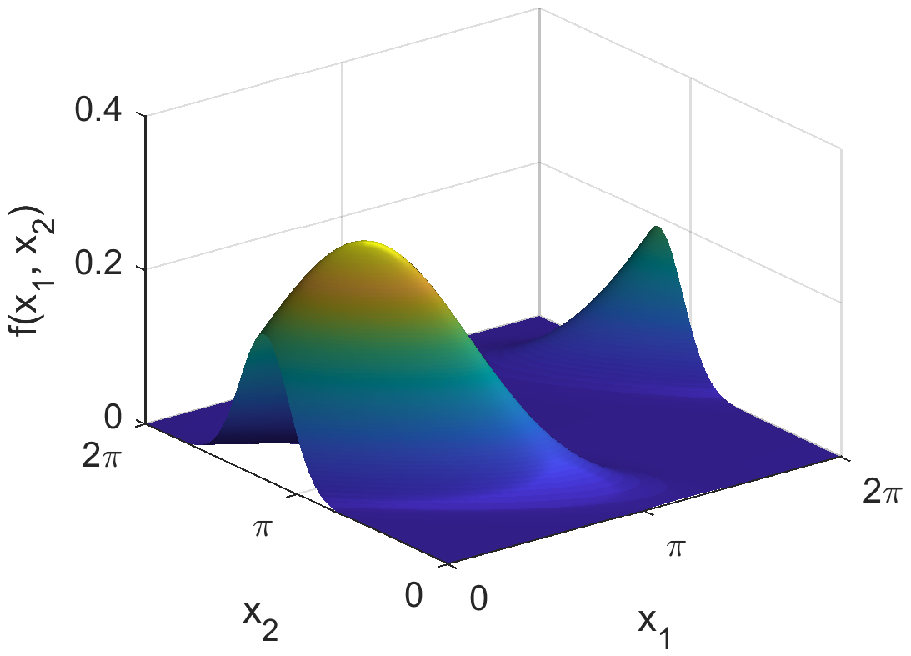}
		\caption{Visualization in the $x_1$--$x_2$--plane. Both $x_1$ and $x_2$ are $2\pi$-periodic.}
		\label{fig:twn}
	\end{subfigure}
	\quad
	\begin{subfigure}{.48\textwidth}
		\centering
		\includegraphics[width=6cm]{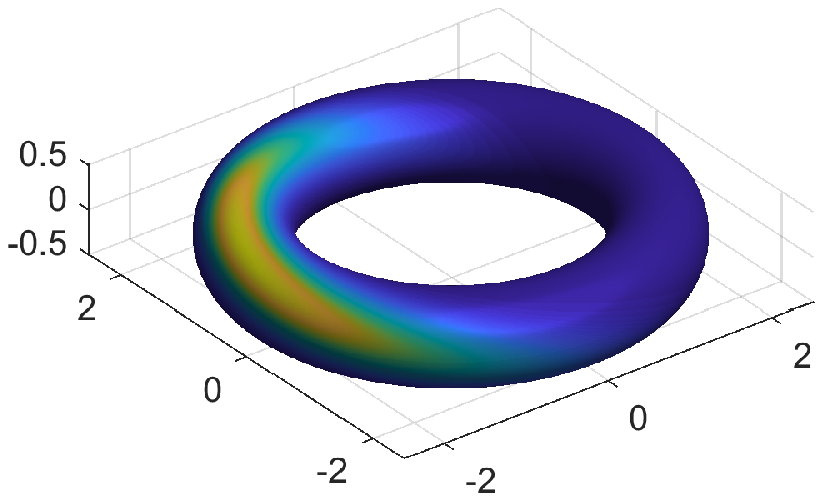}
		\caption{Visualization on the torus. 
		}
		\label{fig:twn-torus}
	\end{subfigure}
	\caption{Density of a bivariate wrapped normal distribution on the torus shown using two different visualizations.}	
	\label{fig:twn-all}
\end{figure}

\begin{example}[Bivariate wrapped normal]
\label{example:toruswn}
First, we instantiate a bivariate wrapped normal distribution using the following statement.
\begin{CodeChunk}
\begin{CodeInput}
> twn = ToroidalWNDistribution([1;3], [1, -0.8; -0.8, 0.9])
\end{CodeInput}
\begin{CodeOutput}
twn = 
  ToroidalWNDistribution with properties:
    dim: 2
     mu: [2x1 double]
      C: [2x2 double]
\end{CodeOutput}
\end{CodeChunk}
We can visualize the density of this distribution as a function $[0,2\pi)^2 \to \mathbb{R}^+$ using the \code{plot} method.
\begin{CodeChunk}
\begin{CodeInput}
> twn.plot();
> setupAxisCircular('x', 'y');
> shading interp; camlight; lighting phong;
> xlabel('x_1'); ylabel('x_2'); zlabel('f(x_1, x_2)');
\end{CodeInput}
\end{CodeChunk}
The resulting plot after adjusting some \proglang{MATLAB} plotting settings is shown in Figure~\ref{fig:twn}. Alternatively, we can create a visualization on the surface of a torus using the \code{plotTorus} method:
\begin{CodeChunk}
\begin{CodeInput}
> twn.plotTorus();
> axis equal; shading interp; camlight; lighting phong;
\end{CodeInput}
\end{CodeChunk}
This yields the plot depicted in  Figure~\ref{fig:twn-torus}. Furthermore, we can investigate the different correlation coefficients for this distribution using the following code.
\begin{CodeChunk}
\begin{CodeInput}
> r1 = twn.circularCorrelationJammalamadaka(), 
  r2 = twn.circularCorrelationJohnson(),
  r3 = twn.circularCorrelationJupp()
\end{CodeInput}
\begin{CodeOutput}
r1 =
   -0.8086
r2 =
   -0.8086
r3 =
   -1.0667
\end{CodeOutput}
\end{CodeChunk}
As you can see, the first correlation coefficient \citep{jammalamadaka1988} and the second correlation coefficient \citep{johnson1977} are identical up to the displayed number of digits in this example. However, this does not hold in general. The value of the coefficient by \cite{jupp1980} is quite different, and is not even restricted to the interval $[-1,1]$. 

It can be shown that for a bivariate wrapped normal distribution, the marginals are wrapped normal. They can be obtained using the following function calls.
\begin{CodeChunk}
\begin{CodeInput}
> wn1 = twn.marginalizeTo1D(1), wn2 = twn.marginalizeTo1D(2)
\end{CodeInput}
\begin{CodeOutput}
wn1 = 
  WNDistribution with properties:
       mu: 1
    sigma: 1
      dim: 1
wn2 = 
  WNDistribution with properties:
       mu: 3
    sigma: 0.9487
      dim: 1
\end{CodeOutput}
\end{CodeChunk}
As can be seen, the marginals are returned as objects of the class \code{WNDistribution}, i.e., a circular distribution that can be used as discussed in Section~\ref{sec:distributions:circle}. Thus, this example illustrates one of the benefits of having implemented distributions on multiple different manifolds within a single library.
\end{example}

\begin{table}[htb]
	\centering
	\begin{tabular}{ll}
		\toprule
		\bf Class name & \bf Comment \\
		\midrule
		CustomHypertoroidalDistribution & allows any user-specified pdf  \\		
		HypertoroidalFourierDistribution  & multivariate Fourier distribution \citep{JAIF16_Pfaff} \\	
		HypertoroidalMixture & mixture of arbitrary hypertoroidal distributions \\		
		HypertoroidalUniformDistribution & uniform distribution on the hypertorus \\
		HypertoroidalVMSineDistribution & multivariate VM, sine version \citep{mardia2008}) \\ 
		HypertoroidalWDDistribution & multivariate wrapped Dirac mixture \citep{kurz2015} \\
		HypertoroidalWNDistribution  & multivariate WN \citep{kurz2015} \\		
		\bottomrule
	\end{tabular}
	\caption{Probability distributions on the hypertorus.}
	\label{table:hypertorusdistributions}
\end{table}

Beyond toroidal distributions, we also offer some hypertoroidal distributions, i.e., distributions on the $n$-torus $[0,2\pi)^n$. An overview of the supported distributions is given in Table~\ref{table:hypertorusdistributions}, all of which are generalizations of the corresponding toroidal distributions.

All hypertoroidal distributions use \code{AbstractHypertoroidalDistribution} as their base class. Because the circle and the torus are special cases of the hypertorus for $n=1$ and $n=2$, respectively, circular and toroidal distributions also inherit (indirectly) from this class. In this way, a lot of code can be shared among many distributions even though they are defined on different manifolds.

\subsection{Real hypersphere}
In this section, we consider probability distributions defined on the unit hypersphere $S^{n-1} = \{\vecx \in \mathbb{R}^n : \Vert \vecx\Vert = 1 \}$, i.e., we parameterize the unit hypersphere as a set of unit vectors in $\mathbb{R}^n$ for some $n\in \mathbb{N}$. Note that this also encompasses the unit circle if $n=2$, but uses a different parameterization (the set of two-dimensional unit vectors rather than a one-dimensional interval of length $2\pi$) compared with the section above.

On the hypersphere, we consider several different distributions as well. The first distribution is the von Mises--Fisher distribution proposed by \cite{fisher1953}, a generalization of the circular von Mises distribution to the unit hypersphere. It is parameterized by a unit vector $\vecmu$ defining the mode of the distribution as well as a concentration parameter $\kappa$ influencing its dispersion. The von Mises--Fisher distribution is unimodal and radially symmetric around $\vecmu$. The Watson distribution \citep{watson1965} is closely related and has the same set of parameters, but it is antipodally symmetric (i.e., $f(\vecx)=f(-\vecx)$), and thus is bimodal with modes at $\pm \vecmu$. However, it is still radially symmetric around the axis of $\vecmu$. In order to represent anisotropic noise, the Watson distribution can be generalized to obtain the Bingham distribution as defined by \cite{bingham1974}. The Bingham distribution is usually parameterized by an orthogonal matrix $\matM$ that defines the location of the mean and the orientation of the principal axes of the uncertainty, as well as a diagonal matrix $\matZ$ responsible for representing the uncertainties along the different axes. Furthermore, we can enforce that the diagonal entries of $\matZ$ are sorted in ascending order and that the last diagonal entry is zero \citep{JAIF14_Kurz-Bingham} without changing the expressiveness of the distribution. Thus, we use this parameterization within \pkg{\libname}\footnote{It should be noted that some authors use slightly different parameterizations, e.g., \cite{glover2014icra}.}.

\begin{table}[bt]
	\centering
	\begin{tabular}{ll}
		\toprule
		\bf Class name & \bf Comment \\
		\midrule
		BinghamDistribution & see \cite{bingham1964}, \cite{bingham1974} \\
		HypersphericalDiracDistribution & discrete distribution on the real hypersphere \\
		HypersphericalUniform & uniform distribution on the real hypersphere \\ 
		VMFDistribution & von Mises--Fisher distribution \citep{fisher1953} \\
		WatsonDistribution & see \cite{watson1965} \\
		\midrule
		BayesianComplexWatsonMixtureModel & complex Watson mixture with prior \\
		ComplexAngularCentralGaussian & see \cite{Kent1997Data} \\
		ComplexBinghamDistribution & see \cite{kent1994} \\
		ComplexWatsonDistribution & see \cite{mardia1999complexwatson} \\
		ComplexWatsonMixtureModel & mixture of complex Watson distributions \\
		\bottomrule
	\end{tabular}
	\caption{Probability distributions on the hypersphere.}
\end{table}

The architecture for hyperspherical probability distributions is similar to that of the previously discussed manifolds. There is a base class called \code{AbstractHypersphericalDistribution}, which provides generic features independent of the particular distribution such as plotting and numerical integration. The individual distributions inherit from this class and can provide methods for the pdf, the normalization constant, stochastic sampling, parameter estimation, etc. As the normalization constant for the Bingham distribution is given by a hypergeometric function of matrix argument, it is quite expensive to evaluate. We have implemented several possible methods including the saddlepoint approximation by \cite{kume2005}. Further discussion about the different methods for computing the normalization constant can be found in \cite{MFI14_Gilitschenski}. Similar to the deterministic sampling schemes on the circle (see Example~\ref{example:detsampling}), we also provide a deterministic sampling scheme for the Bingham distribution, which is presented in \cite{TAC16_Gilitschenski}, and a deterministic sampling scheme for the von Mises--Fisher distribution proposed in \cite{SPL16_Kurz}.




\subsection{Complex hypersphere}
This section introduces three distributions and related statistical models defined on the complex hypersphere $\mathbb C S^{n-1} = \{\vecz \in \mathbb{C}^n : \Vert \vecz \Vert = 1 \}$.
The complex Bingham distribution is defined by its probability density function
\begin{align*}
	p(\vecz; \matB) = \frac{1}{c_{\mathrm B}(\matB)} \exp(\vecz^{\hermite} \matB \vecz), \quad \vecz \in \mathbb C S^{n-1} 
\end{align*}
with Hermitian transpose $\vecz^{\hermite} := \bar{\vecz}^\transpose$, where $\vecz$ is conditioned on $\vecz^{\hermite} \vecz = 1$, $c_{\mathrm B}(\matB)$ is an appropriate normalization term and $\matB$ is the complex positive semi-definite parameter matrix.
The complex Bingham distribution has complex symmetry, namely, it is invariant under scalar rotation, i.e.,  $p(\vecz) = p(\vecz \exp(\complexUnit \varphi))$.
Similar to the real Bingham distribution, its deviation around the mean is governed by the difference between the eigenvalues of $\matB$.
Again, the parameter matrices $\matB$ and $\matB + k\eye$ define the same distribution \citep{kent1994}.
A maximum likelihood fit according to~\cite{kent1994} is implemented in \code{ComplexBinghamDistribution.fit()}.

In contrast to the real case, the complex Bingham normalization constant $c_{\mathrm B}(\matB)$ can be written in terms of elementary functions
\begin{align}
	c_{\mathrm B}(\matB) = 2 \pi^n \sum\limits_{k=1}^{n} a_k \exp \lambda_k \ ,
	\qquad a_k^{-1} = \prod\limits_{k\neq l} (\lambda_k - \lambda_l) \ ,
\end{align}
where $\lambda_k$ are the eigenvalues of the parameter matrix $\matB$.
A symbolic implementation is used to generate code for the normalization constant \code{ComplexBinghamDistribution.logNorm()} and other moments of the complex Bingham distribution.

Counterintuitively, the complex Bingham distribution is a special case of the real Bingham distribution of higher dimension. The corresponding real Bingham parameter matrix can be calculated by replacing each entry $B_{kl}=\alpha_{kl} \exp(\complexUnit \varphi_{kl})$ with blocks $\tilde{B}_{kl}$
\begin{align*}
	\tilde{B}_{kl} = \alpha_{kl} \begin{pmatrix}
		\cos(\varphi_{kl}) & -\sin(\varphi_{kl}) \\ \sin(\varphi_{kl}) & \cos(\varphi_{kl})
	\end{pmatrix}.
\end{align*}
This relationship is useful to test implementations of the complex distributions against their real counterparts and is implemented in \code{ComplexBinghamDistribution.toReal()}. Nevertheless, the algorithms for the complex case can be implemented more efficiently without relying on the real counterpart.

The complex Watson distribution is a special case of the complex Bingham distribution.
It is defined by its pdf \citep{mardia1999complexwatson}
\begin{align*}
p(\vecz; \kappa, \vecw) = \frac{1}{c_{\mathrm W}(\kappa)} \exp(\kappa |\vecz^{\hermite} \vecw|^2), \quad \vecz \in \mathbb C S^{n-1} \ ,
\end{align*}
where $\vecw \in \mathbb C S^{n-1}$ is a complex vector with unit norm and $\kappa$ governs the concentration around the mean direction.
The complex Watson normalization constant $c_{\mathrm W}(\kappa)$ can be written in terms of elementary functions \citep{mardia1999complexwatson}.
An implementation of the normalization constant is provided in \code{ComplexWatsonDistribution.logNorm()} and derivatives thereof are used for maximum likelihood estimates in \code{ComplexWatsonDistribution.fit()}.

Different sampling algorithms for the complex Bingham distribution are known \citep{kent2004}.
Here, a sampling process based on sampling from truncated exponential distributions has been implemented in \code{ComplexBinghamDistribution.sample()}.
The sampling process can be used to create samples for a complex Watson distribution and for complex Watson mixture models.

Early applications of the complex Watson and complex Bingham distributions are statistical modeling of two-dimensional landmarks \citep{kendall1984}.
A fairly recent application is to describe phase and level differences in multi-channel recordings \citep{vu2010,drude2014}.
Different speaker positions cause different phase and level differences such that an EM algorithm for clustering can be employed.
An EM algorithm for a complex Watson mixture model is implemented in \code{ComplexWatsonMixtureModel.fit()} \citep{vu2010}.
An extension to the EM algorithm incorporates prior knowledge about the mode direction and the mixture weights of each mixture component.
The corresponding variational EM algorithm is given by \code{BayesianComplexWatsonMixtureModel.fit()} \citep{drude2014}.

\begin{figure}
	\centering
	\includegraphics[width=6cm]{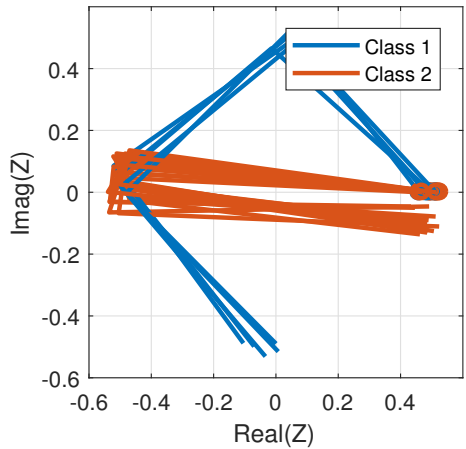}
	\caption{Samples from a complex Watson mixture model in the complex shape domain.}
	\label{fig:cWMM}
\end{figure}

Figure~\ref{fig:cWMM} shows some samples of a complex Watson mixture model in the complex shape domain \citep{kent1994}.
Although pdfs on complex hyperspheres cannot be visualized easily, plots in the shape domain allow visually inspecting similarities between shapes.
The unnormalized mode vectors of the underlying complex Watson distributions are
\begin{align*}
	\vecw_1 = \begin{bmatrix}
	1 & \complexUnit & -1 & -\complexUnit
	\end{bmatrix}^{\transpose}
	\quad\text{and}\quad
	\vecw_2 = \begin{bmatrix}
	1 + 0.1\complexUnit & -1 + 0.1\complexUnit & -1 - 0.1\complexUnit & 1 - 0.1\complexUnit
	\end{bmatrix}^{\transpose},
\end{align*} respectively.

Kent suggested the complex angular central Gaussian model as an alternative to the complex Bingham distribution.
Its probability density function is given by
\begin{align*}
	p(\vecz, \matSigma) = \frac{\Gamma(D)}{2\pi^D} |\matSigma|^{-1} \left( \vecz^{\hermite} \matSigma^{-1} \vecz\right), \quad \vecz \in \mathbb C S^{n-1} \ ,
\end{align*}
where $\Gamma(\cdot)$ refers to the gamma function \citep{Kent1997Data}.
The library provides the probability density function as well as a sampling algorithm.
Additionally, parameter estimation is provided in \code{ComplexAngularCentralGaussian.fit()}.
Natural extensions are a Complex Angular Central Gaussian Mixture Model \citep{ito2016complex} and a Complex Bingham Mixture Model \citep{ito2016modeling}, both of which found great applications in speech enhancement. We plan to add the corresponding code to \pkg{\libname} as future work.
 
\subsection{SE(2)}
Finally, we consider the manifold of rigid body motions in two dimensions called $SE(2)$. A rigid body motion can be seen as a rotation together with a translation. The rotation is an element of the group of two-dimensional rotations $SO(2)$---which can be parameterized as the unit circle---and the translation is a vector in $\mathbb{R}^2$. In \pkg{\libname}, we consider two different continuous distributions on $SE(2)$. As they use different parameterizations of $SE(2)$, the distributions do not share a common base class. 

\subsubsection{Partially wrapped normal distribution on SE(2)}
The first distribution is called the partially wrapped normal distribution (PWN) presented in \cite{MFI14_Kurz} that was further discussed in \citet[Section~2.3.3]{kurz2015}\footnote{\cite{roy2014} refer to this distribution as semi-wrapped Gaussian. In \cite[eq.~(78)]{lo1975}, the term $(n,m)$-folded normal is used.}. This distribution is defined on $[0,2 \pi) \times \mathbb{R}^2$ and is obtained from a normal distribution on $\mathbb{R}^3$ where the first component is wrapped. Then, the first component can be used to represent the angle of the rotation, whereas the second and third components are used to represent the translation. In analogy to the WD distribution on the circle and the bivariate WD distribution on the torus, we also define a partially wrapped Dirac mixture (PWD) distribution. Similar to the toroidal expectation values given in \eqref{eq:toroidalmoments}, it is of interest to consider the moments
\begin{align}
	\label{eq:se2moments}
	\tilde{\vecmu} = \E \left( \begin{bmatrix} \cos(x_1) \\ \sin(x_1) \\ x_2 \\ x_3 \end{bmatrix} \right) , \
	\tilde{\matC} = \E \left( \left( \begin{bmatrix} \cos(x_1) \\ \sin(x_1) \\ x_2 \\ x_3 \end{bmatrix} - \tilde{\vecmu} \right) \cdot \left( \begin{bmatrix} \cos(x_1) \\ \sin(x_1) \\ x_2 \\ x_3 \end{bmatrix} - \tilde{\vecmu} \right)^\transpose \right) \ ,
\end{align}
where $[x_1, x_2, x_3]^\transpose$ is a partially wrapped random variable. Note that although both consider a four-dimensional augmented random vector, \eqref{eq:toroidalmoments} and \eqref{eq:se2moments} differ by their treatment of the last two dimensions. The values defined in \eqref{eq:se2moments} are available using the methods \code{mean4D} and \code{covariance4D}. Moreover, we provide the ability to obtain the marginals of a \code{SE2PWNDistribution} as a \code{WNDistribution} and as a \code{GaussianDistribution}, respectively.

\subsubsection{Modified Bingham distribution}
The second distribution on $SE(2)$ is related to the Bingham distribution in the sense that it also arises by restricting a Gaussian random vector \citep{Fusion14_GilitschenskiKurz}. This is motivated by the fact that a multiplicative subgroup of dual quaternions can be used for representing elements of $SE(2)$, which is reminiscent of the approach by \cite{matsuda2014}. Similar to the Bingham case, our distribution needs to be antipodally symmetric in order to account for the fact that unit dual quaternions are a double cover of $SE(2)$. The distribution is characterized by its pdf
\begin{align*}
	f(\ux) = \fr{1}{N(\fC)} \exp\li(\ux^\transpose\,\fC\,\ux\ri)\ , \qquad\ux\in S^1\times\R^2\ ,
\end{align*}
where  $N(\fC)$ denotes the normalization constant. This corresponds to the density of a four-dimensional Gaussian where the first two entries of $\ux$ are interpreted as one vector that is restricted to unit length. This probability distribution is implemented within the class \code{SE2BinghamDistribution}.

Not every choice of $\fC\in\R^{4\times 4}$ is admissible for this distribution. In order to improve our understanding of the structure of the underlying distribution, we rewrite $\fC$ as 
\begin{align*}
	\fC = \bbmat \fC_1 & \fC_2^\top \\ \fC_2 & \fC_3 \ebmat
\end{align*}
and we can then rewrite the pdf as 
\begin{align*}
f(\ux_s,\ux_t) 
 = N(\fC)^{-1} \cdot \exp\big( \ux_s^\top \fT_1 \ux_s 
    + (\ux_t-\fT_2\ux_s)^\top \fC_3(\ux_t-\fT_2\ux_s)\big)\ ,
\end{align*}
where $\ux_s\in S^1$, $\ux_t\in\R^2$ and $\fT_1=\fC_1-\fC_2^\top \fC_3^{-1}\fC_2$, $\fT_2=-\fC_3^{-1}\fC_2$ with $\fC_1,\,\fC_2,\,\fC_3\in\R^{2\times 2}$. 
From this representation, it follows that $\fC_1$ needs to be symmetric (but not necessarily positive or negative definite), $\fC_2$ may be arbitrary, and $\fC_3$ has to be symmetric negative definite.

An \code{SE2BinghamDistribution} object can be constructed using either of the two constructors \code{SE2BinghamDistribution(C)} or \code{SE2BinghamDistribution(C1,C2,C3)}. 
Besides that, we have implemented a method for obtaining the covariance matrix using Monte Carlo integration (\code{computeCovarianceMCMC}), a computation of the normalization constant (\code{computeNC}), the mode of the density (\code{mode}), a parameter estimation procedure (\code{fit}), and deterministic as well as random sampling procedures (\code{sampleDeterministic} and \code{sample}).


%% file: filters.tex
Based on the probability distributions introduced in the previous section, it is possible to derive recursive filtering algorithms to perform recursive Bayesian estimation. In the following, we introduce a number of filters that are implemented in \pkg{\libname}.

\subsection{Circle}
Several filters for the unit circle are available in \pkg{\libname}. These include both filters based on circular statistics and traditional filters originally intended for linear domains that have been modified for use on the unit circle and that can be employed for comparison. In the following, we distinguish the circular filters based on the type of density they use.

\subsubsection{WN-assumed filter}
First of all, we have implemented several WN-assumed filtering algorithms in the class \code{WNFilter}, i.e., filters approximating the true distribution with a WN distribution after each prediction and filtering step. This class allows predicting with a noisy identity system model, with a nonlinear measurement model in conjunction with additive noise \citep{ACC13_Kurz}, and with a nonlinear measurement model in conjunction with non-additive noise. As far as the measurement model is concerned, the class can handle noisy identity measurements and nonlinear measurements given by a likelihood \citep{ACC14_Kurz} with several methods.  A more thorough discussion of these scenarios can be found in~\cite{AES16_Kurz}. 

\subsubsection{VM-assumed filter}
Analogously, we can assume a von Mises distribution instead of a wrapped normal distribution. This alternative is implemented in \code{VMFilter}. Similar to before, we also distinguish between different types of system and measurement models as discussed in~\cite{AES16_Kurz}. It should be noted that for identity system and identity measurement models, the filter proposed by~\cite{azmani2009} arises as a special case. Finally, we also implement a measurement update based on nonlinear measurement functions (rather than measurement likelihoods) proposed in~\cite{Fusion15_Gilitschenski}.

\subsubsection{Fourier filters}
The Fourier identity filter and the Fourier square root filter, as explained in~\cite{Fusion15_Pfaff,JAIF16_Pfaff}, use a truncated Fourier series to approximate the density or the square root of the density. While this filter is very universal and only makes few assumptions about the noise distributions, nonlinear measurement models require knowledge of the likelihood and for nonlinear system models, the transition density is required. Unless an identity system model with additive noise is used for the prediction step, the transition density needs to be given as one of the toroidal distributions~\citep{Fusion16_Pfaff} depending on the current state and the state at the next time step.
The Fourier filters are particularly powerful as we implemented the ability to approximate other distributions using the Fourier series representation described in Section~\ref{sec:distributions:circle}. Thus, prediction and filter steps can easily be performed in an approximate fashion for arbitrary densities. For this filter, the user only needs to specify at least one additional parameter at the time of initialization. The integer $n$ (also called \code{noOfCoefficients}) must be given to determine the number of Fourier coefficients. The optional string input argument \code{transformation} can be provided to specify if the Fourier identity filter or the Fourier square root filter should be used. If no second argument is given, the square root filter is used by default. Higher values of $n$ result in better approximations for distributions that are more peaked but also yield a moderately higher run time as the filter has an asymptotic complexity of $O(n\log n)$ for the update step and for the prediction step with an identity model with additive noise. If a prediction step using the transition density is required, the run time complexity increases to $O(n^2\log n)$.

\subsubsection{Gaussian-assumed filters}
In order to assess the performance of filters based on directional statistics in comparison with filters making a Gaussian assumption, we included modified versions of the unscented Kalman filter (UKF) \citep{julier2004} based on the implementation of \cite{steinbring2015}. For this purpose, we consider two different possibilities how the filter can be modified as discussed in \cite{AES16_Kurz} and \cite[Section~3.1]{kurz2015}. First, we can define the filter on a (local) chart of the manifold, e.g., the open interval $(0, 2\pi)$ and try to detect and fix issues when the boundary of the chart is reached by reparameterizing if necessary. In this case, the resulting filter has a scalar state. This type of filter is implemented in the class \code{CircularUKF}. Second, it is also possible to consider a filter on the space in which the manifold is embedded and to introduce a constraint enforcing that the state always resides on the manifold. In this case, the resulting filter has a two-dimensional state vector, which is constrained to be of unit length. This type of filter is implemented in the class \code{ConstrainedUKF}.

\subsubsection{Particle filter}
A commonly used filter that avoids the Gaussian assumption is the particle filter \citep{arulampalam2002}. As this filter does not really depend on the underlying manifold as long as the system function and the measurement likelihood properly consider the periodicity, it is easy to adapt the particle filter to the circle. Hence, we have implemented a particle filter with sequential importance resampling (SIR) in \code{CircularParticleFilter}. The particle filter is a Markov Chain Monte Carlo method, and thus, relies on random sampling, so it constitutes a nondeterministic method.

\subsubsection{Discrete filter}
We also consider a Dirac-based discrete filter based on an evenly spaced grid on the circle~\citep{MFI16_Kurz}, which was previously used by \cite{Fusion15_Pfaff}. Similar approaches have been applied to practical problems, e.g., localization of a robot \citep{burgard1996}. The discrete filter closely resembles the particle filter, but it uses equally spaced particles with fixed positions. Consequently, prediction and measurement update only affect the weights of the particles but not their location. The discrete filter is deterministic and can closely approximate the exact Bayesian filter provided a sufficient number of grid points is used.

\subsubsection{Piecewise Constant Filter}
In addition to the Dirac-based discrete filter discussed above, we also offer a filter based on piecewise constant distributions \citep{MFI16_Kurz} that is similar to a Wonham filter \citep{wonham1964}. This filter subdivides the interval $[0,2\pi)$ into a predefined number of smaller intervals of equal size and assumes a uniform distribution within each interval. The filter requires a system matrix that contains the transitions probabilities from each interval into any other interval, which can be precomputed using numerical methods. For the measurement update, it is possible to use a likelihood function or to discretize the measurement space and use a precomputed measurement matrix containing the conditional probabilities of obtaining each discrete measurement given the state is in a certain interval.

\begin{figure}
	\centering
	\begin{subfigure}{.48\textwidth}
		\centering
		\includegraphics[width=65mm]{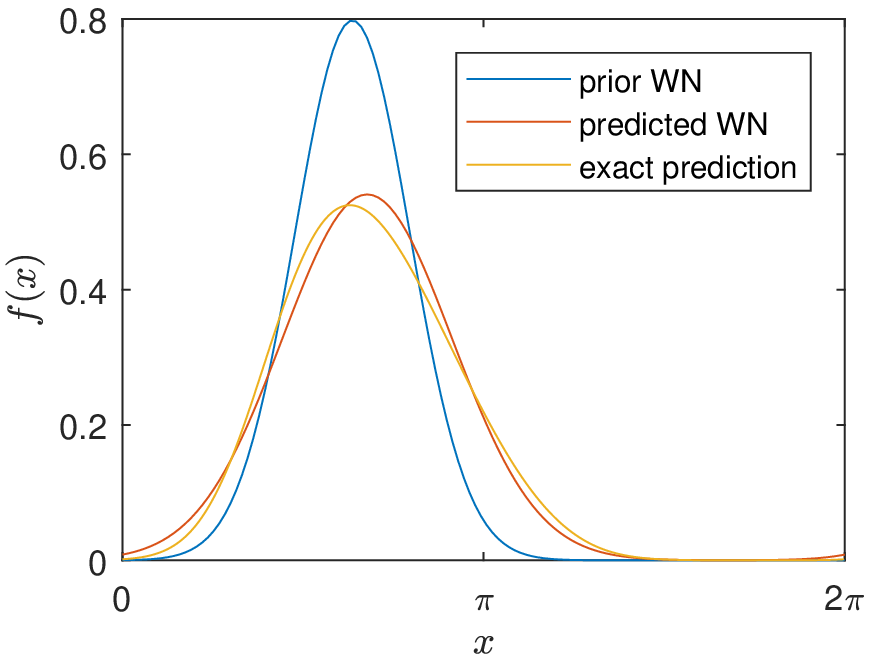}
		\caption{Prediction step.}
		\label{fig:prediction}
	\end{subfigure}
	\begin{subfigure}{.48\textwidth}
		\centering
		\includegraphics[width=65mm]{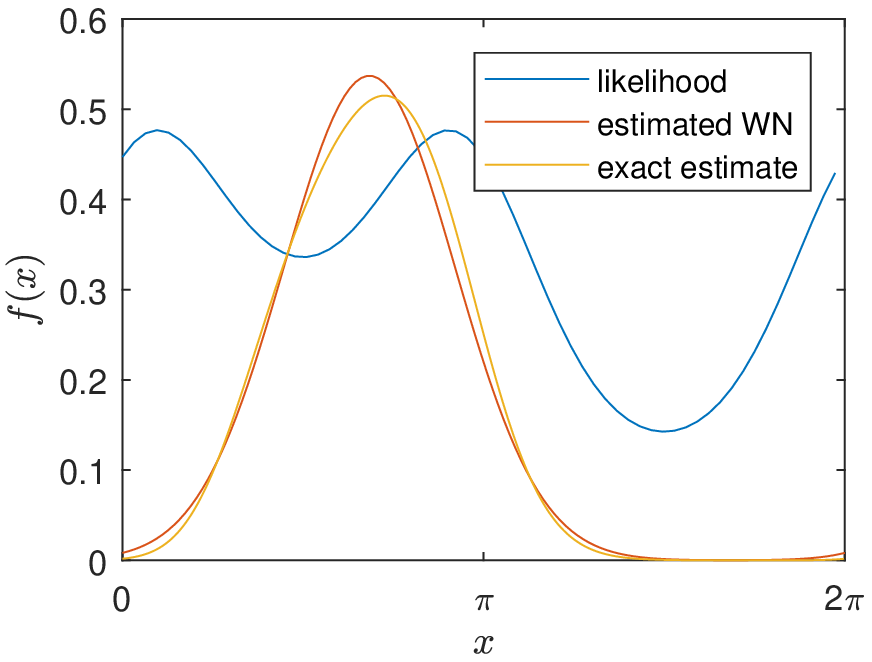}
		\caption{Measurement update step.}
		\label{fig:update}
	\end{subfigure}
	\caption{Results of the nonlinear circular filtering example. We show the result obtained with the WN-assumed filter in comparison with the true result.}	
	\label{fig:filteringexample}
\end{figure}

\begin{example}[Nonlinear circular filtering]
Let us consider a system with a circular state $x_k \in [0, 2\pi)$ and system dynamics
\begin{align*}
	x_{k+1} &= a(x_k) + w_k \ , \\
	a_k(x_k) &= (x_k + 0.5 \cdot \cos^2(x_k)) \mod 2 \pi \ ,
\end{align*}
where $w_k \sim \mathcal{WN}(x; 0, 0.4)$ is WN-distributed additive noise. If we assume that the current state is distributed according to $x_k \sim \mathcal{WN}(x; 2, 0.5)$, we can perform the prediction step with the WN-assumed filter using the following commands.
\begin{CodeChunk}
\begin{CodeInput}
> filter = WNFilter();
> filter.setState(WNDistribution(2, 0.5));
> a = @(x) mod(x + 0.5 * cos(x)^2, 2*pi);
> filter.predictNonlinear(a, WNDistribution(0, 0.4));
> filter.getEstimate()
\end{CodeInput}
\begin{CodeOutput}
ans = 
  WNDistribution with properties:
       mu: 2.1289
    sigma: 0.7377
      dim: 1
\end{CodeOutput}
\end{CodeChunk}
It can be seen that the predicted density is returned as a wrapped normal distribution (see Figure~\ref{fig:prediction}). Now we consider the measurement model 
\begin{align*}
	\hat{z}_k &= h_k(x_k) + v_k 
\end{align*}
with
\begin{align*}
	h_k: [0,2 \pi) \to \mathbb{R} , \  h_k(x_k) &= \sin(x_k) \ ,
\end{align*}
where $v_k \sim \mathcal{N}(x; 0, 0.7)$ is normally distributed additive noise. In this case, we have a circular state, but a real-valued measurement. However, a circular measurement (or a measurement on a completely different manifold) would be possible as well. If we obtain a measurement, say $\hat{z} = 0.3$, we can perform the measurement update as follows.
\begin{CodeChunk}
\begin{CodeInput}
> h = @(x) sin(x);
> measurementNoise = GaussianDistribution(0, 0.7);
> likelihood = LikelihoodFactory.additiveNoiseLikelihood(h, measurementNoise);
> filter.updateNonlinearProgressive(likelihood, 0.3)
> filter.getEstimate()
\end{CodeInput}
\begin{CodeOutput}
ans = 
  WNDistribution with properties:
       mu: 2.1481
    sigma: 0.7427
      dim: 1
\end{CodeOutput}
Once again, we obtain the result as a wrapped normal distribution (see Figure~\ref{fig:update}).
\end{CodeChunk}
\end{example}

\subsection{Torus and hypertorus}
Circular filtering can be generalized to hypertoroidal filtering, just as circular distributions can be generalized to hypertoroidal distributions. As toroidal and hypertoroidal filtering is not a well-researched field, only a few filters are available. In \pkg{\libname}, we provide a filter called \code{ToroidalWNFilter}, which is based on the toroidal wrapped normal distribution proposed in \cite{CDC14_Kurz} along with extensions for nonlinear system and measurement equations given in \cite{Fusion17_Kurz}. Furthermore, there is a modified version of the unscented Kalman filter \citep{julier2004} called \code{ToroidalUKF}, which can be seen as a two-dimensional generalization of the \code{CircularUKF} discussed above. We also provide toroidal and hypertoroidal generalizations of the particle filter in \code{ToroidalParticleFilter} and \code{HypertoroidalParticleFilter}, respectively. The class \code{HypertoroidalFourierFilter} offers a generalization of the Fourier filters used in the circular case. The implementation of additional hypertoroidal filters is planned as future work.

\subsection{Hypersphere}
In \pkg{\libname}, we have also included several filters for estimation on the unit hypersphere. A nonlinear filter based on the von Mises--Fisher distribution \citep{SPL16_Kurz} is available in the class \code{VMFFilter}, which inherits from the base class for hyperspherical filters called \code{AbstractHypersphericalFilter}. The class \code{VMFFilter} includes the filters by \cite{chiuso1998,markovic2014icra} as a special case. A hyperspherical particle filter and a hyperspherical version of the UKF are also available. 

We have also implemented filters assuming antipodal symmetry, i.e., we assume that the points $-\vecx$ and $\vecx$ have the same probability density. This type of filter can be used for axial estimation problems, e.g., when the axis of rotation of an object is to be estimated and the direction of the axis is irrelevant. Furthermore, antipodal symmetry appears in unit quaternions, which can be used for estimation on $SO(3)$. 

All hyperspherical filters with antipodal symmetry are derived from the abstract base class \code{AbstractAxialFilter}. The name refers to the estimation of an axis, e.g., a rotation axis. We provide the Bingham filter described in \cite{JAIF14_Kurz-Bingham}, a special case of which was also considered by \cite{glover2013tr}. Furthermore, the Bingham filter contains the unscented extension proposed in \cite{TAC16_Gilitschenski}. For comparison, \pkg{\libname} also includes an axial version of the Kalman filter \citep{kalman1960} in the class \code{AxialKalmanFilter}. This Kalman filter uses a Gaussian distribution to approximate one of the two modes of the bimodal Bingham distribution (on one of the hemispheres).

\subsection{SE(2)}
We also implemented two different filters for estimation of planar rigid-body motions based on a subgroup of the unit dual quaternions that can be represented as four-dimensional vectors where the first two entries are restricted to unit length (when jointly considered as a two-dimensional vector) as proposed by \cite{Fusion14_GilitschenskiKurz}. One of the filters is based on the modified Bingham distribution (in \code{SE2BinghamFilter}) and is similar to the structure of the Bingham filter  \citep{MFI15_Gilitschenski}. The other filter is a UKF that is implemented (in \code{SE2UKF}) in the same way as the hyperspherical UKF. The only difference is that \code{SE2UKF} ensures the resulting estimate to be a unit dual quaternion.